\documentclass[12pt]{article}
\usepackage[dvips]{graphicx,color}
\newcommand{\dif}{\mathrm{d}}
\begin{document}

\title{Obtaining the large angle MSW solution to the solar neutrino
problem in models}
\author{Ilja Dorsner and S.M. Barr\\
Bartol Research Institute\\
University of Delaware\\ 
Newark, DE 19716}

\date{\today}

\maketitle

\begin{abstract}
The large mixing angle (LMA) MSW solution to the solar neutrino problem 
seems favored by the data at the moment over the small mixing angle (SMA)
MSW solution and the vacuum (VAC) solution. In this paper the various main 
types of models of neutrino masses and mixings are studied from the point of
view of how naturally they can give the LMA solution. Special attention is
given to a very simple type of ``lopsided" $SU(5)$ model. 
\end{abstract}

\newpage

\section{Introduction}

The main solutions to the solar neutrino problem are the SMA solution
(small mixing angle MSW), the LMA solution (large mixing angle MSW),
the LOW solution, and the VAC solution (vacuum oscillations). 
The experimental situation has been and remains very ambiguous.
However, recent results seem
somewhat to disfavor the SMA and VAC solutions. In fact the 
superKamiokande collaboration claims that they
are excluded at the 95\% confidence level \cite{superk}. 

On the other hand, a survey of the hundreds of models of neutrino 
masses and mixings published in the last three years shows that
most of them yield the SMA or VAC solution, and even some that claim
to obtain the LMA solution are only marginally consistent
with the latest global analyses of the data. The purpose of this paper 
is to look at the main types of models of neutrino masses and 
mixing angles that have been proposed in the literature from the point 
of view of their ability to yield the LMA solution in a natural way. 
There are two aspects of this question that can be distinguished.
First, one can ask whether a certain scheme or model can {\it fit} the
LMA solar solution with some choice of model parameters that is not
too badly fine tuned.
Second, one can ask whether the model {\it explains}
the LMA values of the neutrino masses and mixing angles. 
In order to say that a theoretical model really explains
them, something close to the LMA best-fit values
should emerge automatically when the parameters of the theoretical
model take their most ``natural values". If a model parameter that is
{\it a priori} of order one must be set to a value of ten or a tenth
in order to fit the neutrino masses and mixings, one has accomodated 
them but not really
explained them. What our survey will show is that of the great number
of models that now exist in the literature, few can be said to provide
an explanation (in this sense) of the LMA values of $\tan^2 \theta_{sol}$
and $\Delta m^2_{sol}$. 

What are the values that are to be explained?
A recent global analysis of Bahcall, Krastev and Smirnov \cite{bks} 
(published before the SNO results
\cite{sno}) gives as the best fit values for the LMA solution 
$\Delta m^2_{sol} = 4.2 \times 10^{-5}$ eV$^2$, and
$\tan^2 \theta_{sol} \simeq 0.26$. However, the allowed region
is fairly broad. The 90\% confidence-level allowed region given in \cite{bks}
extends in $\Delta m^2_{sol}$ from about $2 \times 10^{-5}$ eV$^2$ to
$10^{-4}$ eV$^2$, and in $\tan^2 \theta_{sol}$ from about 0.15 to 0.5;
i.e. about a factor of two in either direction for both parameters.
A significant aspect of the fits in \cite{bks} is that they nearly exclude
exactly maximal mixing for the LMA solution. 
The 95\% confidence level contour extends
in $\tan^2 \theta_{sol}$ only up to 0.55, and the 99\% contour up
to 0.7. A more recent analysis by Bahcall, Gonzalez-Garcia and Pena-Garay
\cite{bgp}, which includes the SNO results, gives similar results for
the LMA solution; their best-fit value of $\tan^2 \theta_{sol}$ being
about 0.4.
However, the allowed LMA regions of \cite{bgp} extend somewhat 
higher in $\tan^2 \theta_{sol}$ than those obtained in \cite{bks}. 
For example, the fit of Fig. 9 of \cite{bgp}, in which sterile neutrinos are 
excluded {\it a priori}, gives a 95\% confidence-level region that
extends up to about $\tan^2 \theta_{sol} = 0.8$, and a 99\%
confidence-level region that extends slightly past $\tan^2 \theta_{sol} = 1$.

As we shall see, the issue of how close to exact maximal
mixing $\tan^2 \theta_{sol}$ is allowed to be is crucial for deciding whether
several kinds of models can naturally give an acceptable LMA solution. 
In this paper we shall say that a model gives an LMA solution that is
in comfortable agreement with the data if it predicts 
$\tan^2 \theta_{sol} \leq 0.8$. 
It is convenient to express the mass-squared splitting $\Delta m^2_{sol}$
as a fraction, which we shall call $r$, of the mass-squared splitting 
relevant to the atmospheric neutrino oscillations: $r \equiv
\Delta m^2_{sol}/\Delta m^2_{atm}$. The best fit values give $r \simeq 
1.4 \times 10^{-2}$.

The rest of the paper is organized as follows. In section 2, we shall
look at the basic non-see-saw approaches to neutrino mass. 
We shall see that models based on such approaches have been constructed  
which can fit the LMA solution, but for the most part not comfortably, 
either because
$r$ tends to come out too small or because the solar mixing angle
tends to come out too close to maximality. In other words, most of the
non-see-saw models that can fit the LMA solution do not really explain it
in the sense that we have defined. 
In section 3, we look at see-saw models.  Here too, most of the published 
models  give the SMA or VAC solutions. However, we show that there are 
some reasonably simple ``textures" that can reproduce nicely the LMA values 
of the neutrino masses and mixings. However, it remains unclear whether 
these simple textures can arise in simple models.

In section 4, we look at a well-known model that is particularly interesting 
for two reasons: (a) it is very simple in conception, and (b)
it can explain at least the LMA value of the solar neutrino angle, although 
it does not explain the value of the neutrino mass splitting.
It is an $SU(5)$ grand unified model with an abelian family symmetry.
We shall analyze this model in some detail both analytically
and numerically. We shall show how the predictions of this model can
be studied statistically in a completely analytic way by assuming that the
unknown parameters of the model have Gaussian distributions. This method
should be easily aplicable to many other kinds of models.

Section 5 is a brief summary.

\section{Non-see-saw models}

\noindent
{\bf A. Non-see-saw models where $\theta_{atm}$ comes from $M_{\nu}$.}

\vspace{0.2cm}

In non-see-saw models the mass matrix $M_{\nu}$ of
the three light neutrinos is typically generated by new low-energy
physics. It therefore has no relation, or only a very indirect
relation, to the Dirac mass matrices of the charged leptons, the
down quarks, and the up quarks (which matrices we denote $L$, $D$, 
and $U$ respectively). This has the great advantage of making it easy
to explain why the atmospheric neutrino mixing angle is very large
($U_{\mu 3} \equiv \sin \theta_{atm} \cong 0.7$) while the corresponding 
quark mixing is so small
($V_{cb} \cong 0.04$). If the Dirac matrices are assumed to be hierarchical,
then they would naturally give the small mixing angles seen in the quark 
sector. But if $M_{\nu}$ is 
unrelated to the Dirac mass matrices, it could easily have a very different 
form with large off-diagonal elements that gives large mixing angles.  
Non-see-saw models based on this idea are called Type I(1) in \cite{bd}.

The tricky question for this type of model is to explain why 
$\Delta m^2_{sol} \ll \Delta m^2_{atm}$. If the large mixing 
$U_{\mu 3}$ comes from
diagonalizing the 2-3 block of $M_{\nu}$, one would expect that
$m_2$ and $m_3$, the second and third eigenvalues of $M_{\nu}$,
would have similar magnitudes, in which case typically so would 
$\Delta m^2_{sol} = m_2^2 - m_1^2$ and $\Delta m^2_{atm} = m_3^2 - m_2^2$.
The challenge then is to reconcile the hierarchy seen in the eigenvalues of
$M_{\nu}$ with the large atmospheric mixing angle. To do this requires
a special form of $M_{\nu}$. Two special forms have been found 
viable in constructing realistic models, one leads to a so-called 
``inverted hierarchy" $m_1 \cong m_2 \gg m_3$, and the other to the ordinary
hierarchy $m_1 \ll m_2 \ll m_3$. We shall consider these in turn.

\vspace{0.2cm}

\noindent
{\bf Inverted hierarchy models.} 

Inverted hierarchy models have the following special form for $M_{\nu}$:

\begin{equation}
M_{\nu} = \left( \begin{array}{ccc} 
m_{11} & c M  & s M \\ c M  & m_{22} & m_{23} \\
s M & m_{23} & m_{33} \end{array} \right).
\end{equation}

\noindent
Here $c \equiv \cos \theta$ and $s \equiv \sin \theta$, where 
$\theta \sim 1$, and $m_{ij} \ll M$. One can diagonalize this matrix in 
stages, the first step being to rotate by angle $\theta$ in the 2-3
plane, bringing the matrix to the form

\begin{equation}
M'_{\nu} = \left( \begin{array}{ccc} 
m_{11} & M & 0 \\ M & m'_{22} & m'_{23} \\ 0 & m'_{23} & m'_{33}
\end{array} \right).
\end{equation}

\noindent
One sees immediately that $m_1 \cong m_2 \cong M \gg m_3$. 
The mass-squared splitting relevant to atmospheric oscillations is
$\Delta m^2_{atm} \cong M^2$, 
whereas the splitting relevant to solar oscillations is
$\Delta m^2_{sol} \cong 2 (m_{11} + m'_{22}) M$,
which is much smaller, as required by all the viable solar solutions.
The atmospheric angle gets a contribution $\theta \sim 1$ from
diagonalizing the 2-3 block of $M_{\nu}$, so in the absence of some
unlikely cancellation it will be large, as observed. 
But what of the solar
angle? From the fact that the 1-2 block of Eq. (2) has a pseudo-Dirac
form it is apparent that the solar mixing angle will be close to maximal.
Consequently, inverted hierarchy models cannot give the SMA solar solution,
but rather give ``bimaximal" mixing.

The inverted hierarchy form of Eq. (1) can arise in several plausible
ways. One example is the Zee type of model \cite{zeetype}. 
In the Zee model \cite{zee} there is a
singly charged singlet scalar field $h^+$, which is allowed by the
standard model quantum numbers to couple (antisymmetrically) to both
a pair of lepton doublets ($h^+ L_i L_j$) and a pair of Higgs doublets
($h^+ \Phi_a \Phi_b$), assuming that more than one Higgs doublet exists.
If both types of coupling are present,
a conserved lepton number cannot be consistently
assigned to $h^+$, and consequently $\Delta L = 2$ Majorana masses for
the left-handed neutrinos arise at one-loop level. The resulting one-loop
mass matrix has the form 

\begin{equation}
M_{\nu} = \left( \begin{array}{ccc}
0 & a & b \\ a & 0 & c \\ b & c & 0 \end{array}
\right). 
\end{equation}

\noindent
For $c \ll a \sim  b$, this has the desired inverted hierarchy form.

The inverted hierarchy form can also arise in models with an
approximately conserved $L_e - L_{\mu} - L_{\tau}$ lepton number.
If this quantum number is exactly conserved, then only the 12, 21, 13, and
31 elements of $M_{\nu}$ can be nonvanishing. If there are small
violations of $L_e - L_{\mu} - L_{\tau}$ the form in Eq. (1) can result
\cite{Lemt}.

The question of present interest to us is whether the inverted hierarchy 
can give an acceptable LMA solution. To answer this one must look more 
closely at the solar neutrino mixing angle. 
This is given by $\theta_{sol} = \theta_{12}^{\nu}
- \theta_{12}^L$, where the two angles on the right-hand side are the
contributions that come from diagonalizing $M_{\nu}$ and $L$ respectively.
From Eq. (2) it is easily found that $\tan 2 \theta_{12}^{\nu} \cong
2 M/(m'_{22} - m_{11})$, so that $\theta_{12}^{\nu}
\cong  \pi/4 - (m'_{22} - m_{11})/4M$. We have already seen that
$r \equiv \Delta m^2_{sol}/\Delta m^2_{atm} \cong 2(m'_{22} + m_{11})/M$.
Requiring that this be of order $10^{-2}$ as needed for the LMA solution,
and assuming that there are no accidental cancellations, one has
that $\theta_{12}^{\nu} \cong \pi/4 + O(10^{-3})$. If $\theta_{12}^L$ vanished,
this would give $\tan^2 \theta_{sol} = 1 + O(10^{-3})$, which is too
close to maximal mixing to be in comfortable accord with the global
fits. However, one expects that $\theta_{12}^L \sim \sqrt{m_e/m_{\mu}}
\cong 0.07$. This contribution can have any complex phase relative
to the contribution from $M_{\nu}$, and can therefore increase or decrease
$\tan^2 \theta_{sol}$ from unity. If one assumes that $\theta_{12}^L = 0.07$ 
and has a relative minus sign to
$\theta_{12}^{\nu}$, then $\tan^2 \theta_{sol} \cong 0.75$, which is 
consistent with the global LMA fits.
However, one can see that the tendency of inverted hierarchy models
is to give solar mixing that is closer to maximality than to the best-fit
LMA value of $\tan^2 \theta_{sol} \approx 0.4$.
This is one reason
why many of the published inverted hierarchy models claim a better fit to
the VAC solution than to the LMA solution \cite{invhier}.
A significant reduction of the experimental 
upper limits on $\tan^2 \theta_{sol}$
would make the inverted hierarchy idea much less plausible as an
explanation of the LMA solution. For example, a value of
$\tan^2 \theta_{sol} = 0.5$, would imply in the inverted hierarchy
context that $\tan \theta_{12}^L \cong 0.17 \cong 2.5 \sqrt{m_e/m_{\mu}}$,
which would require a very special form of the 1-2 block of $L$.

\vspace{0.2cm}

\noindent
{\bf Ordinary hierarchy models.}

The other possibility for non-see-saw models that gives a large
atmospheric neutrino mixing angle coming from $M_{\nu}$ and a
hierarchy in the mass-squared splittings is

\begin{equation}
M_{\nu} \cong \left( \begin{array}{ccc}
m_{11} & m_{12} & m_{13} \\ m_{12} & s^2 M & sc M \\
m_{13} & sc M & c^2 M \end{array} \right).
\end{equation}

\noindent
Here, again, $c \equiv \cos \theta$ and $s \equiv \sin \theta$, where
$\theta \sim 1$, and $m_{ij} \ll M$. As written, the 2-3 block of the matrix
has vanishing determinant; however, it is assumed that there are small
corrections to these elements, which we have not written.

As in the case of the inverted hierarchy models, one can diagonalize this
in stages, starting with a rotation by angle $\theta$ in the 2-3 plane.
The result of such a rotation is to bring the matrix to the form

\begin{equation}
M'_{\nu} = \left( \begin{array}{ccc}
m_{11} & m'_{12} & m'_{13} \\ m'_{12} & m'_{22} & 0 \\
m'_{13} & 0 & M \end{array} \right).
\end{equation}

\noindent
Because of the small corrections to the 2-3 block that were just
mentioned, the 22 element in Eq. (5) does not vanish, but is small
compared to $M$. This matrix gives $\Delta m^2_{sol} = O(m_{ij}^2)$
and $\Delta m^2_{atm} \cong M^2$. Thus the right hierarchy of splittings
for any of the solutions can be achieved for the appropriate values of
$m_{ij}/M$. In contrast to the inverted hierarchy form,
this form can give either small or large $\theta_{sol}$, and in the
large-angle case there is no preference for values of $\theta_{sol}$
that are very close to maximal. 

The form in Eq. (4) is clearly special in the sense that the 2-3 block is
approximately of rank one. This would be unnatural unless some symmetry
or mechanism guaranteed it. One possibility is that this form arises from a
nonabelian flavor symmetry \cite{mn}, however, this is difficult to
achieve. Rather,
almost all published models that achieve this form in a natural
way use the idea of factorization. The idea of factorization is that
the dominant contribution to the neutrino mass matrix has the form
$(M_{\nu})_{ij} = f_i f_j$, which is obviously of rank one. If
$f_1 \ll f_2, f_3$, this dominant term reproduces the large elements
in Eq. (4). The condition that $f_1$ is small compared to $f_2$ and $f_3$ is
necessary to satisfy the experimental constraint that $U_{e3} \leq 0.15$.
One drawback of most models based on factorization is that they do not
explain why $f_1$ is small.

A factorized form can arise in various ways in non-see-saw models.
A much studied example is supersymmetry with terms in the superpotential
that violate both lepton number and R-parity. Cubic terms of this type
are $\lambda_{ijk}  L_i L_j E^+_k$ and $\lambda'_{ijk} L_i Q_j D^c_k$. 
The latter leads to one-loop $\Delta L = 2$ neutrino mass diagrams,
in which a neutrino converts into a virtual quark-squark pair.
Assuming that the LR squark masses are proportional to the corresponding
quark masses, this diagram gives $(M_{\nu})_{ij} \propto
\lambda_{ikl}^{\prime *} \lambda'_{jlk} m_{d_k} m_{d_l}$. Consequently,
the b-quark/b-squark loop dominates, and gives a contribution that
is proportional to $\lambda_{i33}^{\prime *} \lambda'_{j33} m_b^2$, which
obviously has a factorized form. This gives only the heaviest neutrino mass,
$m_3$. The second largest neutrino mass comes from a similar diagram
with both b and s quarks/squarks in the loop. Consequently, one
has that $r \cong (m_2/m_3)^2 \sim (m_s/m_b)^2 \sim 3 \times 10^{-4}$.
This is much smaller than the value of $1.4 \times 10^{-2}$ 
preferred by experiment;
however, there are several unknown parameters that come into this
calculation, such as the couplings $\lambda'_{ijk}$, so that 
nothing prevents the right LMA value of $r$ from being obtained 
\cite{cubic}. However, the model does not really explain the magnitude
of $r$.

We have only considered the effects of the cubic lepton-number-violating
and R-parity-violating 
terms in the superpotential. There are also in general bilinear terms
of the form $L_i H_u$. These have the effect of mixing leptons and Higgs
fields, and so allow the sneutrino fields to acquire non-vanishing
vacuum expectation values. That, in turn, through the 
sneutrino-neutrino-neutralino 
coupling gives a tree-level neutrino mass in which
the neutralino plays the role of ``right-handed neutrino".
It is easily seen that this tree-level mass has a factorized form
and gives mass only to one neutrino, i.e. $m_3$. The other neutrino
masses, $m_2$ and $m_1$, arise from the one-loop diagrams
previously discussed. In consequence, in such models where both
cubic and bilinear R-parity-violating terms contribute to $M_{\nu}$
one expects that $r \cong (m_2/m_3)^2 \sim (loop/tree)^2
\ll 10^{-2}$. For this reason, most analyses of supersymmetric models
in which the bilinear R-parity-violating terms contribute
to $M_{\nu}$ conclude that there is much more parameter space for the
VAC solution than for the LMA solution, i.e. the LMA solution requires 
special choices or tuning of parameters \cite{susy}. However, in 
\cite{romao} it is shown that under certain assumptions (specifically, 
that there are only bilinear R-parity-violating terms and that
the SUSY-breaking terms are non-universal) the LMA solution can be 
achieved without fine tuning. Nevertheless, it seems, on the whole, that
the SUSY models with R-parity breaking do not do well in explaining
the LMA value of $\Delta m^2_{sol}$.

Another possibility for obtaining an approximately factorized form
for $M_{\nu}$ that has been much studied in the literature 
is called ``single right-handed neutrino dominance" (SRHND) \cite{srhnd}.
As the name suggests, the idea here is that instead of there being
three right-handed neutrinos, one in each family, as there are
in typical grand unified theories or Pati-Salam models, there is just one
right-handed neutrino, $N^c$, which can have mass terms
$M_R N^c N^c + f_i (\nu_i N^c) \langle H \rangle$. Integrating out $N^c$ gives
a rank-1 factorized contribution to $M_{\nu}$. If one
assumes that $N^c$ couples with almost equal strength 
to the $\mu$ and $\tau$ neutrinos, and (for some reason not generally
explained) only weakly to the electron neutrino, the large terms in 
Eq. (4) are reproduced.

One way to explain the smallness of the coupling of $N^c$ to the electron
neutrino would be to impose a symmetry that distinguishes $\nu_e$ from
the $\nu_{\mu}$ and $\nu_{\tau}$ (but does not distinguish the latter
from each other). Such a symmetry would also tend to suppress mixing between 
the $\nu_e$ and the heavier neutrinos, and thus give the SMA solar solution,
as in the model of \cite{mr}.
 
Since the right-handed neutrino only gives mass to one neutrino, some
other mechanism must be found to give mass to the other neutrinos.
In \cite{dk} the lighter two neutrino masses arise from loop effects.
In \cite{st} they arise at tree level from integrating out other 
heavy states that have different flavor quantum numbers than $N^c$.
In \cite{gv} they arise from operators of the form 
$\nu_i \nu_j H_u H_u/M_{Pl}$, which it is argued are generally there anyway,
the idea being to avoid having to invent new beyond-the-standard-model
physics to account for each kind of neutrino mass. In all these cases,
$m_1$ and $m_2$ are much less than $m_3$, though the specific reason is
different in each case: in \cite{dk} they are suppressed by loop factors, 
in \cite{st} by small flavor-breaking parameters, and in \cite{gv} by 
$M_R/M_{Pl}$. That SRHND models tend to give a strong hierarchy in neutrino
masses is what one would naturally expect.
Since the mechanisms that generate the largest neutrino mass 
and the other neutrino masses are different, there is no reason 
{\it a priori} that they should yield masses of similar scale. Rather, 
it would be a coincidence calling for an explanation if they did.
Because most SRHND models give $m_2 \ll m_3$ they yield the VAC solution 
or SMA solution to the solar neutrino problem rather than the LMA
solution \cite{dk, st, gv}. 

To obtain the LMA solution, one wants $m_3$ and $m_2$ to be only about
a factor of ten in ratio. This suggests that they arise from the same
basic mechanism. One possibility is that all three neutrino masses arise 
from integrating out right-handed neutrinos, but that one of those
right-handed neutrinos is somewhat lighter than the others and so
dominates to some extent, but not by a large factor. However, this 
would really be just a special case of the ordinary see-saw mechanism,
which we will discuss in the next section. In fact, the structure given
in Eq. (13) is really based on this idea, which was proposed in \cite{afm}.

\vspace{0.5cm}

\noindent
{\bf B. Non-see-saw models where $\theta_{atm}$ comes from $L$.}

\vspace{0.2cm}

There is another class of non-see-saw models in which the large atmospheric
neutrino angle comes predominantly from the diagonalization of the charged
lepton mass matrix $L$. (This class of models
is called Type II(1) in \cite{bd}.)
This has the advantage that it becomes easy to
reconcile the largeness of $\theta_{atm}$ with the smallness
of $\Delta m^2_{sol}/\Delta m^2_{atm}$, since the former comes from $L$
while the latter comes from $M_{\nu}$. On the other hand another
issue arises for this class of models, namely explaining why the CKM
angles are small.
Since the form of $L$ is such as to give a large mixing angle $\theta_{atm}$,
one would naturally expect that the Dirac mass matrices 
$D$ and $U$ of the quarks would be such as to give similarly large
contributions to the CKM angles. The point is that in many kinds of
models the Dirac mass matrices $L$, $D$, and $U$
are closely related to each other.

One possibility is that there are indeed large contributions to the CKM 
angles coming from $U$ and $D$, but that these nearly cancel. This 
possibility is realized in a much-studied class of models  based
on the idea of ``flavor democracy" \cite{fd}. In flavor democracy
models it is assumed that all the Dirac mass matrices have approximately
the ``democratic" form 

\begin{equation}
\left( \begin{array}{ccc}
1 & 1 & 1 \\ 1 & 1 & 1 \\ 1 & 1 & 1 
\end{array} \right).
\end{equation}

This form can be enforced by permutation symmetries among the three
families.
In the limit of exact flavor democracy, the matrices $U$ and $D$ are
exactly of the same form, so that flavor mixing in the quark sector cancels
out. On the other hand, it is assumed that the neutrino mass matrix
$M_{\nu}$ has a very different form. In most papers
it is assumed to be approximately proportional to the identity matrix,
though in some papers it is only assumed to be nearly diagonal.
As a result, for the leptonic mixing angles there is no cancellation
such as makes the CKM angles small. 

The MNS mixing matrix for the leptons is given by $U_{MNS} =
U_L^{\dag} U_{\nu}$, where $U_L$ and $U_{\nu}$ are the unitary matrices
that diagonalize respectively $L^{\dag} L$ and $M_{\nu}^{\dag} M_{\nu}$.
If $L$ has exactly the democratic form, then 

\begin{equation}
U_L^{\dag} = \left( \begin{array}{ccc}
1/\sqrt{2} & -1/\sqrt{2} & 0 \\ 1/\sqrt{6} & 1/\sqrt{6} & -2/\sqrt{6} \\
1/\sqrt{3} & 1/\sqrt{3} & 1/\sqrt{3} \end{array}
\right) \equiv U_{FD}.
\end{equation}

\noindent
If the mass matrix of the neutrinos is exactly diagonal, then 
$U_{MNS} = U_L^{\dag} = U_{FD}$. This would give 
$\sin^2 2 \theta_{atm} = 8/9$, which is consistent with the data,
and $\tan^2 \theta_{sol} = 1$, i.e. exactly maximal mixing for
solar neutrinos. However, the matrix $L$ clearly cannot have exactly
the democratic form, as that is rank one and would give
$m_e = m_{\mu} = 0$. There must therefore be small corrections to $L$ coming
from the breaking of the permutation symmetries. These corrections not only
generate masses for the electron and muon but also make the angle 
$\theta_{sol}$
deviate from maximality. For the simplest and most widely assumed
form of these corrections to $L$, 
one can calculate the corrections
to $\theta_{sol}$ in terms of $\sqrt{m_e/m_{\mu}}$. One finds, still
assuming that $M_{\nu}$ is diagonal, that \[\tan^2 \theta_{sol}
= 1 - \frac{4}{\sqrt{3}} \sqrt{m_e/m_{\mu}} \cong 0.84,\] 
\noindent
or equivalently
$\sin^2 2 \theta_{sol} = 0.993$. This is too close to unity to be
in comfortable agreement with the LMA global fits. Almost
all published models based on flavor democracy have $\tan^2 \theta_{sol} 
\cong 1$, or else obtain smaller values by fine-tuning.
However, Tanimoto, Watari, and Yanagida have a version in which 
there are small corrections to $M_{\nu}$ that can
reduce $\tan^2 \theta_{sol}$ to the region preferred by the LMA fits
\cite{twy}. While this shows that it is possible within the flavor democracy 
framework to construct LMA models that can fit the data, it does not appear
that flavor democracy does a good job of explaining the LMA value of 
$\tan^2 \theta_{sol}$. Flavor democracy is more naturally
compatible with the VAC or LOW solutions.

We may summarize the situation by saying that most schemes that have been
proposed based on non-see-saw mechanisms neither very comfortably
fit nor really do much to explain the values of the neutrino parameters 
required for the LMA solution to the solar neutrino problem. 
The great majority of non-see-saw models in the literature more
naturally give the SMA or VAC solutions. There are exceptions, which we
have noted above. How close $\tan^2 \theta_{sol}$ is to 1 is a crucial
issue.

\section{See-saw models}

The see-saw mechanism is usually associated with grand unification.
In $SO(10)$ grand unified models, and in most other unified schemes
except $SU(5)$, the existence of one right-handed neutrino 
for each family is required to make up complete
multiplets of the unified group. Moreover, the see-saw formula
$M_{\nu} = - N^T M_R^{-1} N$, where $M_R$ is the Majorana mass matrix
of the right-handed neutrinos and $N$ is the Dirac mass matrix of the
neutrinos, gives neutrino masses in the range required by experiment if
the scale of $M_R$ is near the grand unified scale. Thus both the
existence of neutrino masses and their magnitude are elegantly accounted
for by the related ideas of grand unification and the see-saw mechanism.
In this section, we shall therefore assume that we are dealing with
a grand unified model.

\vspace{0.5cm}

\noindent
{\bf A. See-saw models where $\theta_{atm}$ comes from $M_{\nu}$.}

\vspace{0.2cm}

In models based on $SO(10)$, there is generally a close relationship
among the four Dirac mass matrices $N$, $U$, $D$, and $L$.
Indeed, in the minimal $SO(10)$ model (which is too simple to be
realistic) $N = U \propto D = L$. The smallness of the CKM angles
and the small interfamily mass ratios of the quarks can be explained
by assuming that the matrices $U$ and $D$ are ``hierarchical"
in form. There are two simple kinds of hierarchical mass matrix that
are frequently encountered in models

\begin{equation}
\left( \begin{array}{ccc} (\epsilon')^2 & \epsilon \epsilon' & \epsilon' \\
\epsilon \epsilon' & \epsilon^2 & \epsilon \\ \epsilon' & \epsilon
& 1 \end{array} \right), \;\;\;\;  
\left( \begin{array}{ccc}
\epsilon' & \epsilon' & \epsilon' \\ \epsilon' & \epsilon & \epsilon 
\\ \epsilon' & \epsilon & 1 \end{array} \right),
\end{equation}

\noindent
where $\epsilon' \ll \epsilon \ll 1$. The entries in these matrices 
as written are to be understood as giving only the order in the small 
parameters of the entries.
The first form in Eq. (8) has what may be called a
``geometric hierarchy", since an off-diagonal element is of
the same order as the geometric mean of the corresponding diagonal
elements. The second form in Eq. (8) has what may be called a
``cascade hierarchy", since the matrix is made up of successive
tiers, a 1-by-1, a 2-by-2, and a 3-by-3, of ever smaller magnitude.
Both forms in Eq. (8), if applied to the quark masses, give
$V_{cb} \sim \epsilon$, $V_{us} \sim \epsilon'/\epsilon$, and 
$V_{ub} \sim \epsilon' \sim V_{us} V_{cb}$. 

While there is as a rule a close relation among the four Dirac mass 
matrices in $SO(10)$, the Majorana mass matrix $M_R$ of the right-handed
neutrinos can be quite different in form. For example, in minimal $SO(10)$
the Dirac mass matrices all come from the same term,
${\bf 16}_i {\bf 16}_j {\bf 10}_H$, whereas the matrix $M_R$
comes from different terms, either
${\bf 16}_i {\bf 16}_j \overline{{\bf 126}}_H$
or ${\bf 16}_i {\bf 16}_j \overline{{\bf 16}}_H \overline{{\bf 16}}_H$.
A reasonable hypothesis is that the CKM angles are small because of the
hierarchical nature of the Dirac matrices, while the largeness of
$\theta_{atm}$ and possibly of $\theta_{sol}$ has to do with the
very different form of $M_R$. Models based on this idea were classified in 
\cite{bd} as Type I(2). 

A potential difficulty with this idea is that if
the Dirac mass matrix of the neutrinos $N$ has a hierarchical form it tends, 
through the see-saw formula, to make $M_{\nu}$ also have a hierarchical
form, indeed a more strongly hierarchical form. For example,
suppose $N = {\rm diag}(\epsilon', \epsilon, 1)$, and 
we parametrize $M_R^{-1}$ as $(M_R^{-1})_{ij} = a_{ij}$. Then

\begin{equation}
M_{\nu} = \left( \begin{array}{ccc}
(\epsilon')^2 a_{11} & \epsilon \epsilon' a_{12} & \epsilon' a_{13} \\
\epsilon \epsilon' a_{12} & \epsilon^2 a_{22} & \epsilon a_{23} \\
\epsilon' a_{13} & \epsilon a_{23} & a_{33} \end{array} \right).
\end{equation}

\noindent
If all the $a_{ij}$ are of the same order, then
$\tan 2 \theta_{23}^{\nu} \cong 2 \epsilon a_{23}/(a_{33} - 
\epsilon^2 a_{22}) \sim \epsilon$. In order for $\theta_{atm}$
to come primarily from diagonalizing $M_{\nu}$, one must have
$\theta_{23}^{\nu} \sim 1$. Clearly, this is only possible with some
special form of $M_R$. One possibility is that $a_{23}/a_{33}
\sim \epsilon^{-1}$. If this is true, then there is a hierarchy
among the elements of $M_R$ that is related to the hierarchy
among the elements of $N$. That is, the atmospheric neutrino mixing
angle is of order unity because of a conspiracy between the
Majorana and Dirac neutrino mass matrices. This would appear to be
somewhat unnatural in a theory in which $M_R$ and $N$ have different
origins, as is typically the case in unified models. On the other hand,
this ``Dirac-Majorana conspiracy" might not be unnatural in a model in
which the same flavor symmetry, and the same small parameter
characterizing the breaking of that symmetry, controlled the structure
of both matrices. A good example of such ``correlated hierarchies"
is the model of \cite{stech}. 

A very important question is whether $\theta_{atm}$ can naturally
be of order unity even if the Dirac matrices are hierarchical and
the parameters in $M_R$ have no direct relationship to those of $N$.
The answer is yes. In \cite{js} and \cite{afm} interesting examples were
found that satisfy these criteria. The specific forms given in
those papers happen to lead to the SMA solar solution, 
but with some modifications they can also yield a satisfactory LMA solution,
as we will now see.

\vspace{0.2cm}

\noindent
{\bf Example 1:}
The following structure is closely related to that in \cite{js}:

\begin{equation}
N = \left( \begin{array}{ccc} d \epsilon' & e \epsilon' & f \epsilon' 
\\ g \epsilon' & a \epsilon & b \epsilon 
\\ h \epsilon' & c \epsilon & 1 \end{array} \right) m_N,
\;\;\;\;
M_R = \left( \begin{array}{ccc} 0 & 0 & A \\ 0 & 1 & 0 \\ A & 0 & 0 
\end{array} \right) m_R.
\end{equation}

\noindent
Here $a$, ..., $h$ are of order one, $\epsilon' \ll \epsilon \ll 1$, and
$(\epsilon'/\epsilon) \epsilon^{-1} \ll A \ll \epsilon^{-1}$.
Keeping only the significant terms, the resulting light neutrino mass 
matrix $M_{\nu} = - N^T M_R^{-1} N$ is

\begin{equation}
M_{\nu} \cong -\left(
\begin{array}{ccc}
O(\epsilon^{\prime 2}) & O(\epsilon \epsilon') & d \epsilon'/A \\
O(\epsilon \epsilon') & a^2 \epsilon^2 & ab \epsilon^2 + e \epsilon'/A \\
d \epsilon'/A & ab \epsilon^2 + e \epsilon'/A & b^2 \epsilon^2 + 
2 f \epsilon'/A \end{array} \right) \frac{m_N^2}{m_R}.
\end{equation}

\noindent
A rotation in the 23 plane by angle $\theta \cong \tan^{-1} (a/b) \sim 1$ 
diagonalizes the 2-3 block and brings the matrix to the form

\begin{equation}
M'_{\nu} \cong - \left( \begin{array}{ccc}
O(\epsilon^{\prime 2}) & -\frac{ad}{\sqrt{a^2 + b^2}} \epsilon'/A &
\frac{bd}{\sqrt{a^2 + b^2}} \epsilon'/A \\
-\frac{ad}{\sqrt{a^2 + b^2}} \epsilon'/A & \frac{2a(af - be)}{a^2 + b^2}
\epsilon'/A & 0 \\ 
\frac{bd}{\sqrt{a^2 + b^2}} \epsilon'/A & 0 & (a^2 + b^2) \epsilon^2 
\end{array} \right) \frac{m_N^2}{m_R}.
\end{equation}

\noindent
If one assumes that $\epsilon'/A \sim \epsilon^2/10$, then it is apparent that
$\Delta m^2_{sol}/\Delta m^2_{atm} \sim 10^{-2}$ as required for the LMA
solution. It is to be observed that the 12 and 21 elements of this 
matrix are of the same order as the 22 element. This is just what is
needed to get the right value of $\theta_{sol}$ for the LMA solution,
i.e. a value that is of order unity, but
{\it not} very close to maximal. For example, if the 12 element
is {\it exactly} equal to the 22 element, 
then $\tan^2 \theta_{sol} \cong 0.39$, which is in excellent agreement
with the LMA best-fit value given in \cite{bgp}. An examination of this
matrix reveals that in obtaining the LMA solution a crucial role
is played by the ``cascade hierarchy" form of $N$. In particular, it
is important that $d$ be of the same order as $e$ and $f$, which would not
be the case if $N$ had a ``geometric hierarchy" form. It should also
be noted that the largeness of the atmospheric angle is also traceable to
the cascade hierarchy form of $N$, and specifically to the fact that
$b$ is of the same order as $a$. 

\vspace{0.2cm}

\noindent
{\bf Example 2:}
The following structure is closely related to that given in \cite{afm}

\begin{equation}
N = \left( \begin{array}{ccc} d \epsilon' & e \epsilon' & f \epsilon' 
\\ g \epsilon' & a \epsilon & b \epsilon 
\\ h \epsilon' & c \epsilon & 1 \end{array} \right) m_N,
\;\;\;\;
M_R = \left( \begin{array}{ccc} B & 0 & 0 \\ 0 & A & 0 \\ 0 & 0 & 1 
\end{array} \right) m_R.
\end{equation}

\noindent
As in the last example, $a$, ..., $h$ are of order one, and 
$\epsilon' \ll \epsilon \ll 1$. If one also assumes that
$\epsilon^2/A \gg \epsilon^{\prime 2}/B \gg 1$, then the light neutrino
mass matrix $M_{\nu} = - N^T M_R^{-1} N$ takes the form (keeping 
only the important terms):

\[
M_{\nu} \cong
\left( \begin{array}{ccc}
O(\epsilon^{\prime 2}/A) & ga \epsilon \epsilon'/A + de \epsilon^{\prime 2}/B  
& gb \epsilon \epsilon'/A + df \epsilon^{\prime 2}/B   \\
ga \epsilon \epsilon'/A + de \epsilon^{\prime 2}/B   & 
a^2 \epsilon^2/A + e^2 \epsilon^{\prime 2}/B &
ab \epsilon^2/A + ef \epsilon^{\prime 2}/B \\
gb \epsilon \epsilon'/A + df \epsilon^{\prime 2}/B   &
ab \epsilon^2/A + ef \epsilon^{\prime 2}/B &
b^2 \epsilon^2/A + f^2 \epsilon^{\prime 2}/B \end{array} \right)
\frac{m_N^2}{m_R}.
\]

\noindent
A rotation in the 23 plane by angle $\theta \cong \tan^{-1} (a/b) \sim 1$ 
diagonalizes the 2-3 block and brings the matrix to the form

\[
M'_{\nu} \cong \left( \begin{array}{ccc}
O(\epsilon^{\prime 2}/A) & \frac{d(be-af)}{\sqrt{a^2 + b^2}} 
\epsilon^{\prime 2}/B & O(\epsilon \epsilon'/A) +
O(\epsilon^{\prime 2}/B) \\
\frac{d(be-af)}{\sqrt{a^2 + b^2}} 
\epsilon^{\prime 2}/B & \frac{(be-af)^2}{a^2 + b^2} \epsilon^{\prime 2}/B
& 0 \\ O(\epsilon \epsilon'/A)
+ O(\epsilon^{\prime 2}/B) & 0 & (a^2 + b^2) \epsilon^2/A
\end{array} \right) \frac{m_N^2}{m_R}.
\]

\noindent
The same remarks apply as in the previous example. The largeness
of both the atmospheric neutrino angle and the solar neutrino angle can be
traced to the ``cascade hierarchy" form of $N$. Because the 12, 21, and 22 
element are of the same order, the solar angle is (as required for the LMA
solution)
of order one, but {\it not} very close to maximal. The right ratio
of mass splittings for the LMA solution can be obtained if
$\epsilon^{\prime 2}/B \sim 10^{-1} \epsilon^2/A$.

These two examples show that there are reasonable forms or ``textures"
for the mass matrices in the context of the see-saw mechanism that
can quite naturally yield the LMA solution. However, actual detailed models
based on these textures have not been constructed. It is also not clear
how simple it is for the seemingly required
``cascade hierarchy" form to arise in the framework
of grand unified models. 
Finally, it should be noted that while some forms 
for $N$ and $M_R$ can
be identified which would naturally give the LMA solution, most of the viable
forms give the SMA or VAC solutions, and indeed the great
majority of see-saw models published in the literature give the
latter solutions rather than the LMA solution.

\vspace{0.5cm}

\noindent
{\bf B. See-saw models where $\theta_{atm}$ comes largely from $L$.}

\vspace{0.2cm}

As we saw in part B. of section 2, there are advantages to models in
which the large atmospheric angle comes primarily from the charged
lepton mass matrix $L$. In particular it becomes easy to
reconcile the largeness of $\theta_{atm}$ with the smallness
of $\Delta m^2_{sol}/\Delta m^2_{atm}$, since they come from different
matrices: the former from $L$ and the latter from $M_{\nu}$. 
As we also saw, however, having a large angle arise from the diagonalization
of $L$ raises the question of why a large CKM angle does not arise
from the diagonalization of the quark mass matrices $D$ and $U$.
The answer in the ``flavor democracy" models, was that the CKM angles
are small by a cancellation caused by an approximate symmetry.
The possibility of a different and very elegant answer arises in
the context of grand unified models, especially if an $SU(5)$
symmetry plays a role in the form of the fermion mass matrices.

In $SU(5)$ the left-handed (right-handed) charged leptons are in the 
same multiplets as the (CP conjugates) of the right-handed (left-handed)
down quarks. Since the large mixing angle $\theta_{atm}$ is a mixing of
left-handed leptons, and specifically of left-handed charged leptons in the 
scenarios we are presently considering, it would generally be 
related by $SU(5)$
to a large mixing angle for the right-handed down quarks. But such
a right-handed mixing angle has nothing to do with the observed CKM angles.
On the other hand, the small CKM angles are related by $SU(5)$ to
small mixings of the right-handed leptons, which are irrelevant to
neutrino oscillation phenomena. 

These considerations lead naturally to the idea that the matrix $L$ is such
as to give large left-handed mixings and small right-handed mixings,
so that $\theta_{atm}$ can be large while $V_{cb}$ is small.
In other words, $L$ must be highly asymmetric or ``lopsided" to use
the term suggested in \cite{abb}. (It should be noted that in $SU(5)$, 
$L$ is related to $D^T$, but
not in general to $U$ or $N$. Thus while the lopsidedness of $L$ 
entails the lopsidedness of $D$, there is no reason to expect $N$
and $U$ to be lopsided, and in the examples we give below they are not.)

Many models have been proposed
based on this idea of lopsided mass matrices \cite{lopsided, loplma}.
These are classified as Type II(2) in \cite{bd}.
The great majority of these models give the SMA solution or the VAC 
solution to the solar neutrino problem. However, it is possible to obtain the
LMA solution as well \cite{loplma}. 
In a lopsided model in which the LMA, LOW, or VAC
solution arises, the large atmospheric angle can come primarily
from the matrix $L$ while the large solar angle can come from the 
matrix $M_{\nu}$. This actually has the virtue of simplicity, since
the form of $M_{\nu}$ is less constrained than in models where 
it must give rise both to a large $\theta_{atm}$ and large $\theta_{sol}$.

An example of how the LMA solution might arise in a see-saw model
with lopsided $L$ is provided by the following matrices:

\begin{equation}
N = \left( \begin{array}{ccc} d \epsilon' & e \epsilon' & f \epsilon' 
\\ g \epsilon' & a \epsilon & b \epsilon 
\\ h \epsilon' & c \epsilon & 1 \end{array} \right) m_N,
\;\;\;\;
M_R = \left( \begin{array}{ccc} 0 & A & 0 \\ A & 0 & 0 \\ 0 & 0 & 1 
\end{array} \right) m_R.
\end{equation}

\noindent
As before, it is assumed that $a$, ..., $h$ are of order one, and that
$\epsilon' \ll \epsilon \ll 1$. Suppose the value of $A$ is such that
$\epsilon^2, \; \epsilon' \ll \epsilon \epsilon'/A \ll 1$.
Then, keeping only the most important terms, the light neutrino mass
matrix has the form:

\[
M_{\nu} \cong - \left( \begin{array}{ccc}
O(\epsilon^{\prime 2}/A) & ad \epsilon \epsilon'/A & bd \epsilon \epsilon'/A 
\\ ad \epsilon \epsilon'/A & 2ae \epsilon \epsilon'/A & 
(af + be) \epsilon \epsilon'/A + c \epsilon \\
bd \epsilon \epsilon'/A & (af + be) \epsilon \epsilon'/A + c \epsilon &
1 \end{array} \right) \frac{m_N^2}{m_R}.
\]

\noindent
Here the 23 and 32 elements are small compared to the 33 element,
leading to a small contribution to $\theta_{atm}$; but that is 
alright, since $\theta_{atm}$ is supposed to arise primarily from
diagonalizing $L$ in this class of models. As in the previous
examples, one sees that the 12, 21 and 22 
elements are of the same order, giving a large, but not nearly
maximal, value of $\theta_{sol}$ as required by the LMA solution.
To get the right ratio of neutrino mass-squared splittings one 
needs $\epsilon \epsilon'/A \sim 10^{-1}$.

\section{An $SU(5)$ pattern}

A particularly interesting kind of pattern can arise very simply
in the context of $SU(5)$ with abelian flavor symmetry.
Consider an $SU(5)$ model with a $U(1)$ flavor symmetry under which the
quark and lepton multiplets have the following 
charge assignments: ${\bf 10}_1 (2)$, ${\bf 10}_2 (1)$, ${\bf 10}_3 (0)$,
$\overline{{\bf 5}}_1 (1)$, $\overline{{\bf 5}}_2 (0)$, 
$\overline{{\bf 5}}_3 (0)$, Let the breaking of the $U(1)$ flavor
symmetry be done by a field $\chi$ having $U(1)$ charge -1, and an
expectation value $\langle \chi \rangle/M_{flavor} = \epsilon \ll 1$.
Then the mass matrices of the quarks and charged leptons will have the
following forms.

\begin{equation}
\begin{array}{ll}
D \sim \left( \begin{array}{ccc} \epsilon^3 & \epsilon^2 & \epsilon \\
\epsilon^2 & \epsilon & 1 \\ \epsilon^2 & \epsilon & 1 
\end{array} \right) m_D, &
U \sim \left( \begin{array}{ccc}
\epsilon^4 & \epsilon^3 & \epsilon^2 \\
\epsilon^3 & \epsilon^2 & \epsilon \\
\epsilon^2 & \epsilon & 1 \end{array} \right) m_U, \\ \\
L \sim \left( \begin{array}{ccc} \epsilon^3 & \epsilon^2 & \epsilon^2 \\
\epsilon^2 & \epsilon & \epsilon \\ \epsilon & 1 & 1 
\end{array} \right) m_D. & 
\end{array}
\end{equation}

\noindent
Note that $L$ and $D$ have the lopsided form. It has been pointed
out in several papers in the literature that these forms give a very
good account of the mass ratios and mixing angles of the quarks and 
leptons \cite{by}.

One can see that the quantities $m_{\mu}/m_{\tau}\simeq 1/17$, $m_s/m_b
\simeq 1/50$, and $V_{cb} \simeq 1/25$ all are of order $\epsilon$.
Thus, $\epsilon$ is roughly of order $1/20$. A consistent value of
$\epsilon$ is obtained from the fact that $m_c/m_t \simeq 1/400$, 
$m_u/m_c \simeq 1/200$, $m_e/m_{\mu} \simeq 1/200$, and $V_{ub}
\simeq 1/300$ are all of order $\epsilon^2$. From the
Cabibbo mixing and the ratio $m_d/m_s$, one would get the somewhat
larger value $\epsilon \sim 1/5$.

The light neutrino mass matrix $M_{\nu}$ arises from the see-saw mechanism;
so to know this matrix exactly
it would be necessary to know $M_R$. However, to
know merely the order in $\epsilon$ of the elements of $M_{\nu}$ it is
not necessary to know the $U(1)$ family charges of the
right-handed neutrinos at all, 
since the effective mass term in which $M_{\nu}$ 
appears involves only the left-handed lepton doublets, which are
in the $\overline{{\bf 5}}_i$. Knowing the $U(1)$ charges of the
$\overline{{\bf 5}}_i$ tells us that

\begin{equation}
M_{\nu} \sim \left( \begin{array}{ccc}
\epsilon^2 & \epsilon & \epsilon \\
\epsilon & 1 & 1 \\
\epsilon & 1 & 1 \end{array} \right) m_D^2/m_R.
\end{equation}

From the forms of $L$ and $M_{\nu}$ it is obvious that the mixing 
$U_{\mu 3}$ of the second and third family neutrinos will get $O(1)$
contributions from both these matrices, thus explaining the largeness
of the atmospheric neutrino mixing angle. Let us imagine now
diagonalizing the 2-3 block of $M_{\nu}$ to get 

\begin{equation}
M'_{\nu} \sim \left( \begin{array}{ccc}
\epsilon^2 & \epsilon & \epsilon \\
\epsilon & m_{2(0)} & 0 \\
\epsilon & 0 & 1 \end{array} \right) m_D^2/m_R.
\end{equation}

\noindent
The entry $m_{2(0)}$ would naturally
be expected to be of order 1. However, for the ratio
$r \equiv \Delta m^2_{sol}/\Delta m^2_{atm}$ to come out to be of
order $10^{-2}$, as required by the LMA solution, $m_{2(0)}$ should
rather be of order 1/10. If we accept this rather mild fine-tuning,
and assume that
$m_{2(0)} \sim 1/10$, something interesting can be observed, namely
that the 12 and 21 elements of $M'_{\nu}$ are of the same order as the 22
element, since $\epsilon \sim 1/20$. Recall that this is just what is
needed for $\tan^2 \theta_{sol}$ to come out to be near the best-fit
LMA value of about 0.3 or 0.4.

This model, then, naturally explains both the value of $\theta_{atm}$
and the LMA value of $\theta_{sol}$, provided that $r \equiv
\Delta m^2_{sol}/\Delta m^2_{atm}$ is set to the LMA value.

Let us now imagine diagonalizing $M'_{\nu}$. The rotation needed to
eliminate the 13 and 31 elements will give a contribution to $U_{e3}$
that is of order $\epsilon$, quite consistent with the present
experimental limit of $0.15$. This leaves the diagonalization of the 1-2
block. In doing this one may neglect the 11 element, since it is of
order $\epsilon^2$. One then finds the simple relations 
(a) $\tan 2 \theta_{sol} \sim 2 \epsilon/m_{2(0)}$, and (b)
$m_2 = m_{2(0)}/(1 - \tan^2 \theta_{sol})$. (Here we have ignored the
$O(\epsilon)$ contribution to $\theta_{sol}$ coming from diagonalizing $L$,
since we are interested in large values of $\theta_{sol}$.) 
From these relations one
can infer roughly what region this model gives in the
standard $\log(\tan^2 \theta_{sol})$--$\log(\Delta m^2_{sol})$ plot.
One sees from (a) that $\tan^2 \theta_{sol} \sim \epsilon^2
(m_{2(0)})^{-2}$,
and from (b) that $\Delta m^2_{sol} \sim (m_{2(0)})^2$. In other
words, in the standard plot the region coresponding to this model
lies roughly along a line with slope $-1$ going through the LMA
allowed region. We shall see shortly, both by a much more careful
analytic calculation and by a Monte Carlo numerical calculation,
that this conclusion is correct. The form of (b) tells us something
else that is interesting. As the value of the solar angle approaches
maximality, i.e. $\tan^2 \theta_{sol} \longrightarrow 1$, the denominator
in (b) approaches zero. Therefore, to maintain a finite value of 
$\Delta m^2_{sol}$ the value of $m_{2(0)}$ must be tuned to be 
extremely small. Thus, one expects the region of greatest probability
in this model
to fade away as $\tan^2 \theta_{sol}$ approaches 1. This is confirmed by
the analytic and Monte Carlo calculations, as we shall see. 

We shall now study the predictions of this model in a statistical way, much
in the spirit of \cite{murayama}. Similar statistical analyses have been done
in several recent papers \cite{stat}, and our results are consistent
insofar as they can be compared with theirs. However, our analysis 
differs in some
respects. We do not treat $\epsilon$ as a free parameter, and seek
to find its optimal value for the various solar solutions. Rather,
we fix $\epsilon$ to the value that best reproduces the mass ratio
$m_{\mu}/m_{\tau}$ and then derive the full region of the $\tan^2 \theta_{sol}$
--$\Delta m^2_{sol}$ plane which results. We also show that by treating
the random variables as having a Gaussian distribution the statistical
predictions of the model can be obtained analytically. We also carry
out a numerical simulation using a non-Gaussian distribution similar
to those used in previous analyses and show that it agrees remarkably well
with the analytic results obtained using a Gaussian distribution. 

To carry out
the statistical analysis we parametrize the neutrino and charged
lepton mass matrices as follows:

\begin{equation}
M_{\nu} = \left( \begin{array}{ccc}
f \epsilon^2 & d \epsilon/\sqrt{2} & e \epsilon/\sqrt{2} \\
d \epsilon/\sqrt{2} & b & c /\sqrt{2} \\
e \epsilon/\sqrt{2} & c /\sqrt{2} & a \end{array} \right) m_D^2/m_R,
\end{equation}

\noindent
and

\begin{equation}
L = \left( \begin{array}{ccc}
O(\epsilon^3) & O(\epsilon^2) & O(\epsilon^2) \\
O(\epsilon^2) & D \epsilon & C \epsilon \\
O(\epsilon) & B & A \end{array} \right) m_D .
\end{equation}

\noindent
We will take the unknown order-one parameters $a$, $b$, $c$, $d$, $e$, $f$, $A$, $B$,
$C$, and $D$, to be complex random variables whose real and imaginary
parts have Gaussian distributions with standard deviation $\sigma$.
For example, if $a = |a| e^{i \theta_a}$, then $P(a)\,\dif a =
(2 \pi \sigma^2)^{-1} \exp(- |a|^2/2\sigma^2) |a| \, \dif |a| \, \dif \theta_a$.
It should be noted that we have put factors of $1/\sqrt{2}$ in 
the off-diagonal elements of $M_{\nu}$. This is the appropriate
normalization to use for a symmetric matrix. 

What we want to calculate is the probability distribution
$P(r,t) \, \dif r \, \dif t$, where $r \equiv \Delta m^2_{sol}/\Delta m^2_{atm}$, as before,
and $t \equiv \tan^2 \theta_{sol}$, given that the order-one unknown parameters
in the mass matrices have Gaussian distributions as described.
We will first describe and give the results of an analytic
calculation of $P(r,t)$, and then present the results
of a Monte Carlo numerical calculation of $P(r,t)$.

A very important point in what follows is that if one does unitary
changes of basis 

\[\begin{array}{ccc}
\left( \begin{array}{c} \nu_{L2} \\ \nu_{L3} \end{array}
\right) \longrightarrow V
\left( \begin{array}{c} \nu_{L2} \\ \nu_{L3} \end{array}
\right)& \mbox{or} &
\left( \begin{array}{c} \ell^-_{L2} \\ \ell^-_{L3} \end{array}
\right) \longrightarrow V 
\left( \begin{array}{c} \ell^-_{L2} \\ \ell^-_{L3} \end{array}
\right),
\end{array}\] 

\noindent
the resulting parameters $a'$, ..., $e'$, $A'$, ..., 
$D'$, have exactly the same Gaussian distributions as the
parameters in the original basis. (This would not be true without
the factors of $1/\sqrt{2}$ in $M_{\nu}$.) This is one fact that
makes the analytic calculation tractable using Gaussian distributions.
Moreover this basis independence is more consistent with the 
group-theoretical approach advocated in \cite{murayama}.

The first thing to do is diagonalize $L$. For our purposes, we need only
diagonalize the 2-3 block to find $m_{\mu}/m_{\tau}$ and the contribution
of $L$ to $\theta_{atm}$.
This involves multiplying the 2-3 block of $L$ from the right 
(which in our convention is
a transformation on the {\it left}-handed leptons) by a unitary matrix
\[U^{[23]}_{\ell} \cong \left( \begin{array}{cc} A & B^* \\ -B & A^* \end{array}
\right) (|A|^2 + |B|^2)^{-1/2}.\]
\noindent
This eliminates the large element $B$ 
in $L$ and makes the 33 element become $\sqrt{|A|^2 + |B|^2}$. 
The new 22 and 23 elements, which can be written
$D' \epsilon$ and $C' \epsilon$ respectively,
have the same Gaussian distribution as do $A$, $B$, $C$, and $D$.
Consequently, one has that \[(m_{\mu})_{rms}/(m_{\tau})_{rms} =
\epsilon |D'|_{rms}/(\sqrt{|A|^2 + |B|^2})_{rms} = \epsilon/\sqrt{2}.\]
\noindent
Thus, the most reasonable value to choose for the small parameter, from
the point of view of lepton physics, is
$\epsilon/\sqrt{2} = m_{\mu}/m_{\tau}$.

The first constraint that we shall impose is that the atmospheric neutrino
mixing comes out to be very close to maximal, as found experimentally.
This angle gets contributions
from the diagonalizations of both $L$ and $M_{\nu}$. It would seem, then,
that we must, in computing $P(r,t)$, take into account the random variables
in both mass matrices. However, a great simplification occurs
because of the use of Gaussian distributions and the resulting basis
independence of the probability distributions. A little thought shows
that one can compute $P(r,t)$ in a basis where the contribution to the 
atmospheric neutrino mixing coming from $L$ has some fixed value, 
and the result will not depend on that value. Thus the parameters in $L$ are
irrelevant to $P(r,t)$. It is simplest in practice to choose
the basis where the entire atmospheric neutrino mixing comes from $L$.
A further simplification is achieved by neglecting
the parameters $e$ and $f$. The parameter
$e$ comes into calculating $U_{e3}$ (which is predicted to be of order
$\epsilon$, and so consistent with the experimental bound
$|U_{e3}| \leq 0.15$), but has a negligible effect on $\theta_{sol}$,
$\theta_{atm}$, and the neutrino masses. The parameter $f$ is
multiplied by $\epsilon^2$, and so is negligible also.
Finally, one can choose a basis where the parameters $a$, $b$, and $d$
are real. That means that the only parameters that come into the calculation
of $P(r,t)$ are $|a|$, $|b|$, $|c|$, $|d|$, and $\theta_c \equiv \arg c$. 
From now on, we shall drop the absolute value signs and denote $|a|$, for
example, simply by $a$.

One begins, then, with a matrix

\begin{equation}
M_{\nu} = \left( \begin{array}{ccc}
0 & d \epsilon/\sqrt{2} & 0 \\
d \epsilon/\sqrt{2} & b & c e^{i \theta_c}/\sqrt{2} \\
0 & c e^{i \theta_c}/\sqrt{2} & a \end{array} \right) m_D^2/m_R,
\end{equation}

\noindent
and a probability distribution

\begin{equation}
P(a,b,c,\theta_c, d) = \frac{a b c d}{2 \pi \sigma^8} 
\,\mathrm{e}^{-\frac{1}{2 \sigma^2}(a^2 + b^2 + c^2 + d^2)}.
\end{equation}

The first step is to diagonalize the 1-2 block of the matrix given in Eq. 
(20), which gives $\tan 2 \theta_{sol} \equiv s = \sqrt{2} d \epsilon/b$.  
This allows the elimination of the random variable $d$ in favor of the 
measureable parameter $s$ or equivalently $t$. 
The 11 and 22 elements of the 
matrix then become $\frac{1}{2}(1 - \sqrt{1 + s^2})b$ and
$\frac{1}{2}(1 + \sqrt{1 + s^2})b$ respectively. The latter quantity we 
shall denote as $b'$. 
The next step is to impose the atmospheric angle constraint. Since we
are working in a basis where the contribution to this angle from
$M_{\nu}$ is vanishingly small, the imposition of this constraint
sets the parameter $c$ to zero. More precisely, if one requires that
the (complex) contribution to the atmospheric angle from $M_{\nu}$ have 
magnitude bounded by some arbitrary small cutoff $\Delta_{atm} \ll 1$, 
the condition on $c$
becomes $(c/\sqrt{2})/(a-b') \leq \Delta_{atm}$. This means that the
integration over $\dif c \, \dif \theta_c$ in the probability 
distribution can be
done, yielding $\int c \, \dif c \, \dif \theta_c = 
\pi (\sqrt{2} (a-b') \Delta_{atm})^2$.
The only remaining variables are then $a$, $b$, and $s$.
The random variable $a$ can be eliminated in favor of the
measurable ratio $r$ of mass-squared splittings using the 
relation $r = (b^2 \sqrt{1 + s^2})/(a^2 - b^{\prime 2})$.
It is a very good approximation here to replace $b'$ by $b$, since
for the whole region of interest either $s$ or $r$ is very small
as we shall see.
After all these steps, one is left with a probability distribution
$P(r,b,s)$. The final step is simply to integrate over the random
variable $b$. Since this integral is a Gaussian it is easily done.
The final result is 

\begin{equation}
P(r,s) \, \dif r \, \dif s = N \, \frac{rs \, \dif r \, \dif s}{1+s^2}
\frac{\displaystyle \left[ 
\sqrt{1 + \frac{\displaystyle r}{\displaystyle \sqrt{1 + s^2}}} - 
\sqrt{\frac{\displaystyle r}{\displaystyle \sqrt{1 + s^2}}} \, 
\right]^2}{\displaystyle \left[ 1 + 
\frac{\displaystyle 2r}{\displaystyle \sqrt{1 + s^2}}
+ \frac{\displaystyle r s^2}{\displaystyle 2 \epsilon^2 
\sqrt{1 + s^2}} \right]^{4}},
\end{equation}

\noindent
where $N$ is just a normalization constant.
Changing variable from $s \equiv \tan 2 \theta_{sol}$ to $t \equiv
\tan^2 \theta_{sol}$ one finds that

\begin{equation}
P(r,t)\, \dif r \, \dif t = N \, \frac{2r \, \dif r \, \dif t}{1-t^2}
\frac{\displaystyle \left[ \sqrt{1 + r \left( \frac{\displaystyle 1-t}{\displaystyle 1+t} \right)} - 
\sqrt{ r \left( \frac{\displaystyle 1-t}{\displaystyle 1+t} \right)} \, \right]^2}{\displaystyle \left[ 1 
+ 2r \left( \frac{\displaystyle 1-t}{\displaystyle 1+t} \right) + 
\frac{\displaystyle 2 r t}{\displaystyle \epsilon^2 (1 - t^2)} \right]^{4}}.
\end{equation}

One can see the qualitative behavior of this function rather easily.
The crucial term is the one containing $\epsilon^2$ in the denominator.
This term forces the product $rt$ to be of order $\epsilon^2$.
This is consistent with what we argued above, namely that the
region of greatest probability in this model has $rt \sim$ constant, i.e.
a line of slope -1 in the $\log(\tan^2 \theta_{sol})$--$\log(\Delta m^2_{sol})$
plane.
Moreover, we see that as $t \longrightarrow 1$, the product $rt$
is forced to be less than or of order $\epsilon^2(1 - t^2) \longrightarrow
0$, so that the probability is suppressed for $t \cong 1$.

\begin{figure}
\begin{center}
\includegraphics[width=4.5in]{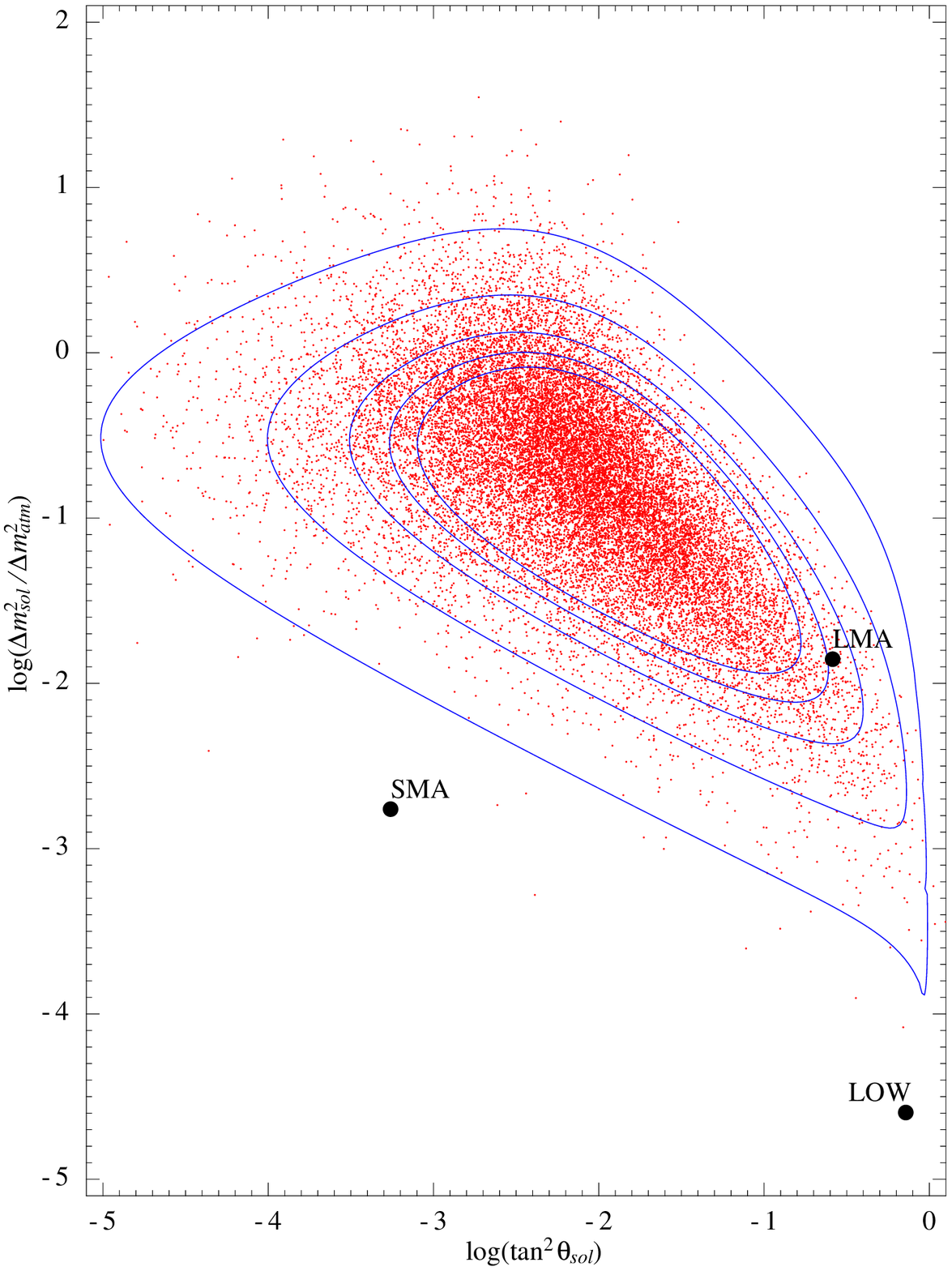}
\end{center}
\caption{\label{FIRST} Contour plot of the normalized probability distribution
$P$ in $\log(\tan^2 \theta_{sol})$--$\log(\Delta m^2_{sol}/\Delta m^2_{atm})$ plane 
with the contour values 0.002, 0.02, 0.06, 0.1, and 0.14 superimposed on
the numerically generated distribution of points. Large dots represent 
best-fit values for LMA, SMA, and LOW solar neutrino solutions, as indicated.}
\end{figure}

In Figure \ref{FIRST}, we give a contour plot of the probability function
just computed analytically, and compare it to the results of a
Monte Carlo calculation. For the Monte Carlo calculation
we used the forms in Eqs. (18) and (19), with $\epsilon/\sqrt{2} =
m_{\mu}/m_{\tau}$, but assumed that the magnitudes 
of the complex random variables, instead of having a Gaussian distribution, 
had constant probability in the interval 0.5 to 2.0 and zero probability outside
that interval. The phases of the complex variables were also treated as random 
numbers and were varied from 0 to $2\pi$. We then diagonalized randomly 
generated matrices to obtain the corresponding MNS mixing matrices $U_{MNS}$ 
and neutrino masses and analyzed the results by imposing the conditions 
$\sin^2 2 \theta_{atm} \ge 0.9$ and $|U_{e3}| \le 0.15$. These conditions 
reduced our initial set of 50,000 data points to 20,860 that were compatible
with both the CHOOZ and the atmospheric neutrino experiments. The points that 
passed the cuts are given in Fig. \ref{FIRST}.

One can see from the excellent agreement between the analytic and Monte Carlo 
results evident in Fig. \ref{FIRST} that the exact form used for the 
probability distributions of the random variables themselves makes little 
difference. This has also been found in other papers \cite{murayama, stat}.
One point that should be noted is that since Figure \ref{FIRST} is a log-log
plot, the correct thing to plot and what has been plotted, is 
$P(\log r, \log t) \sim P(r,t) r t$. 

\begin{figure}
\begin{center}
\includegraphics[width=5in]{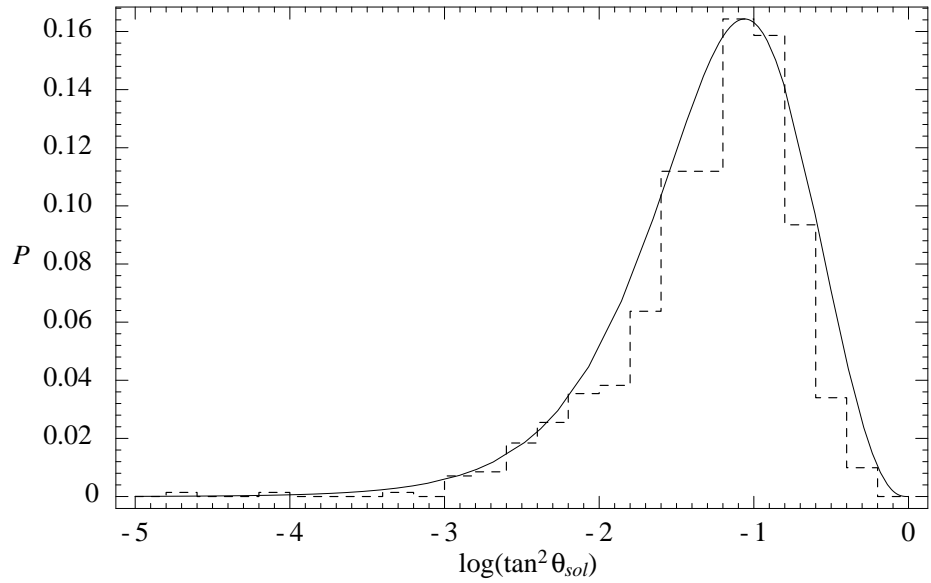}
\caption{\label{SECOND} Normalized probability distribution $P$ (solid line)  as a function of
$\log(\tan^2 \theta_{sol})$ for best-fit LMA solution value $\Delta m^2_{sol}/\Delta m^2_{atm} = 1.4 \times 10^{-2}$ 
plotted against the normalized, binned distribution (dashed line).}
\end{center}
\end{figure}

In Figure \ref{SECOND}, we have taken the slice $r = 1.4 \times 10^{-2}$, which
comes from using the best-fit values from experiment, and plotted
the resulting $P(\log t)$ against a normalized, binned distribution. The binned distribution 
has been obtained by counting the number of data points in the strip 
$\log r = \log(1.4 \times 10^{-2}) \pm 0.1$, the width of one bin being 0.2, and  
the normalization has been carried out with respect to the maximum of $P(\log t)$.
Note that the most probable value for $\tan^2 \theta_{sol}$ is about 
0.1, with a very substantial part of the area under the curve being in the
region [0.2, 0.8] preferred by the LMA solution global fits.

The one weakness of this model is that it does not explain why
$r \equiv \Delta m^2_{sol}/\Delta m^2_{atm} \approx 1.4 \times 10^{-2}$. 
From Figure \ref{FIRST} one sees that a value of $10^{-1}$ for this ratio
is near the peak of the probability distribution $P(r,t)$. However,
the same figure shows that a value of $1.4 \times 10^{-2}$ is near the
edge of the preferred region, and so requires a mild fine-tuning.
However, once $r$ is constrained to be the right value, the
value of $\tan^2 \theta_{sol}$ needed for the LMA solution emerges
quite naturally, as can be seen from Fig. \ref{SECOND}. The 
atmospheric mixing angle is, of course, also naturally explained. 
In Figure \ref{FIRST} the best fit values of the LMA, SMA, and LOW
solutions are indicated by dots. One sees that this $SU(5)$ lopsided
model (with $\epsilon/\sqrt{2} = m_{\mu}/m_{\tau}$) naturally prefers 
the LMA solution over the others.

\section{Conclusions}

One can see from the foregoing that it is considerably easier to build
satisfactory models of the VAC, LOW, or SMA type than of the LMA type.
That is reflected in the models that have actually been constructed
in the literature.
One problem is that in many models which predict large solar mixing angle, 
notably inverted hierarchy schemes and flavor democracy schemes, 
this angle tends to come out very close to maximal. They do not naturally
explain why $\tan^2 \theta_{sol}$ should come out in the range
0.2 to 0.8 preferred by the data. Other non-see-saw schemes, such as
SUSY with R-parity breaking and single-right-handed-neutrino-dominance
(SRHND) models, tend to predict a value of
$\Delta m^2_{sol}/\Delta m^2_{atm}$ significantly less than
that preferred by the LMA solution. While the LMA value of this ratio can be
fit, it is not really explained. 

The situation seems more promising for the see-saw approaches, although here
also the great majority of published models give the small angle or
vacuum solar solutions. We showed that certain fairly simple textures exist 
that would naturally reproduce the neutrino masses and
mixings required by the LMA solution. Whether these textures can be 
implemented in simple models remains to be seen.

One of the few existing schemes that shows a natural
preference for the LMA solution is the lopsided $SU(5)$ model studied in
Section 4. The value of $\Delta m^2_{sol}/\Delta m^2_{atm}$ requires a
mild fine-tuning, but given that, both the atmospheric angle and the
LMA value of the solar angle emerge quite naturally.
We studied the predictions of this model in a statistical way, and found
that by using Gaussian distributions the analysis could be carried out
very simply and accurately by purely analytic means. We believe that the
same methods should be applicable to many other models. The advantage
of such statistical analyses is that they allow one to estimate in a
somewhat objective and quantitative way how "fine tuned" models must be
to reproduce the data.

\end{document}